\documentclass[12pt]{iopart}
\usepackage{graphicx}
\usepackage{amssymb}\usepackage{ulem}
\begin{document}
\title{Equilibrium constants of Hydrogen and Helium isotopes at low nuclear densities.}

\author{R. Bougault$^1$, 
E. Bonnet$^2$, 
B. Borderie$^3$, 
A. Chbihi$^4$, 
J. D. Frankland$^4$, 
E. Galichet$^{3,5}$,
D. Gruyer$^1$, 
M. Henri$^4$, 
M. La Commara$^6$, 
N. Le Neindre$^1$, 
I. Lombardo$^7$, 
O. Lopez$^1$, 
L. Manduci$^{1,8}$, 
M. Parl\^og$^1$, 
R. Roy$^9$, 
G. Verde$^7$, 
M. Vigilante$^6$}
\address{$^1$ Normandie Univ, ENSICAEN, UNICAEN, CNRS/IN2P3, LPC Caen, F-14000 Caen, France}
\address{$^2$ SUBATECH UMR 6457, IMT Atlantique, Universit\'e de Nantes, CNRS-IN2P3, 44300 Nantes, France}
\address{$^3$ Institut de Physique Nucl\'eaire, CNRS/IN2P3, Univ. Paris-Sud, Universit\'e Paris-Saclay, F-91406 Orsay cedex, France}
\address{$^4$ Grand Acc\'el\'erateur National d'Ions Lourds (GANIL), CEA/DRF-CNRS/IN2P3, Bvd. Henri Becquerel, 14076 Caen, France}
\address{$^5$ Conservatoire National des Arts et Metiers, F-75141 Paris Cedex 03, France}
\address{$^6$ Dipartimento di Fisica 'E. Pancini' and Sezione INFN, Universit\`a di Napoli 'Federico II', I-80126 Napoli, Italy}
\address{$^7$ INFN - Sezione Catania, via Santa Sofia 64, 95123 Catania, Italy}
\address{$^8$ Ecole des Applications Militaires de l'Energie Atomique, BP 19 50115, Cherbourg Arm\'ees, France}
\address{$^9$ Laboratoire de Physique Nucl\'eaire, Universit\'e Laval, Qu\'ebec, Canada G1K 7P4}
\begin{abstract}
Equilibrium constants for Hydrogen and Helium isotopes as a function of density and temperature are measured in the framework of the study made by Qin et al. \cite{QinPRL108}. We review and comment on all stages of the analysis and conclude that our measurements are not inconsistent with Qin et al. results. Improvements are being made to the initial analysis and we raise the issue of the binding energies which has to be clarified.  
\end{abstract}

\submitto{\jpg}
\maketitle
\ack{This work is part of the INDRA collaboration program. We thank the GANIL staff for providing us the beams and for the technical support during the experiment. We acknowledge support from R\'egion Normandie under RIN/FIDNEOS.}
\section{Introduction}
In a recent article, Qin et al. \cite{QinPRL108} have investigated clustering 
in low density nuclear matter. Equilibrium constants for $^{2}$H, $^{3}$H,
$^{3}$He and $^{4}$He have been measured following Guldberg and Waage law \cite{PaponLeblond} by selecting experimentally a sub-set of events 
corresponding to a gas of neutrons and protons
in equilibrium with clusters.\par 
Besides the fact that these results are important for the knowledge of the density dependence of the nuclear equation of state \cite{BaldoProgPartNuclPhys}, the measurements are related to density and temperature values that are of major importance for astrophysics since
nuclear equation of state plays a fundamental role in the understanding of core-collapse supernovae, mergers of compact stars and cooling proto-neutron stars for example.
In particular the chemical composition of proto-neutron stars influences the neutrino opacities and then their cooling \cite{HempelPRC91} \cite{PaisPRC97}.\par 
This result therefore deserves to be confirmed or disproved using another experiment and equipment.\par
This paper presents the analysis framework of Qin et al. \cite{QinPRL108} adapted to our data concerning H and He isotopes. Then we raise the problem of cluster binding energy values which are assumed to be the vacuum values for the determination of the density which seems inconsistent to us.\par
In our experiment the use of four entrance channel systems with different neutron to proton ratios (N/Z) at the same bombarding energy, $^{136,124}$Xe+$^{124,112}$Sn, will allow to test different assumptions made during the analysis.       
\section{Experimental details}
The $4\pi$ multi-detector INDRA 
\cite{Pou95} was used to study four reactions with beams of $^{136}$Xe and $^{124}$Xe, 
accelerated by the GANIL cyclotrons to 32 MeV/nucleon, and thin (530 $\mu$g/cm$^{2}$) targets of $^{124}$Sn and $^{112}$Sn. 
INDRA is a charged product multidetector,
composed of 336 detection cells arranged in 17 rings centered
on the beam axis and covering 90\% of the solid angle. 
The first ring (2$^{o}$ to 3$^{o}$) is made of 12 telescopes composed of
300 $\mu$m silicon wafer (Si) and CsI(Tl) scintillator (14 cm
thick). Rings 2 to 9 (3$^{o}$ to 45$^{o}$) are composed of 12 or 24
three-member detection telescopes : a 5 cm thick ionization
chamber (IC); a 300 $\mu$m or 150 $\mu$m silicon
wafer; and a CsI(Tl) scintillator (14 to 10 cm thick) coupled
to a photomultiplier tube. 
Rings 10 to 17 (45$^{o}$ to 176$^{o}$) are composed of 24, 16 or 8 two-member telescopes:
an ionization chamber and a CsI(Tl) scintillator of 8, 6
or 5 cm thickness.
INDRA can identify in charge fragments from Hydrogen to Uranium and in mass light fragments with low thresholds.
Recorded event functionality was activated under a triggering factor based on a minimum number of fired 
telescopes (M$_{trigger}^{min}$) over the detector acceptance (90\% of $4\pi$). 
During the experiment 
minimum bias (M$_{trigger}^{min}$=1) and exclusive (M$_{trigger}^{min}$=4) data were recorded.\par
This study is limited to the forward part of the centre of mass (hereinafter called c.m.) 
for which excellent mass and charge identification performances are achieved for Hydrogen and Helium isotopes (hereinafter called lcp). Since only most violent collisions will be used in the following, only exclusive data are analysed.
\section{Event and sub-event selections}
In this paragraph, we present the event (central collisions) and the sub-event (c.m. angular cut) selections used in order to study a gas of neutrons and protons in equilibrium with clusters.\par  
In a previous publication \cite{BougaultPRC97} concerning the same experiment, we have demonstrated that chemical equilibrium is achieved in central collisions. Therefore the most violent events, as in \cite{QinPRL108}, are employed in order to look for sub-events that correspond to a gas of neutrons and protons in equilibrium with clusters. Events with reduced impact parameters lower than 0.15 have been selected, using the impact parameter evaluator described in \cite{BougaultPRC97} (forward c.m. lcp total transverse energy greater than 200 MeV). As will be explained later, the gas of particles is assumed to have an isotropic momentum distribution. Since very central collisions have the advantage of minimizing the sideflow \cite{ReisdorfAnnRevNucl1997}, the centrality selection is thus appropriate.\par
In \cite{BougaultPRC97} we have also indicated for the studied systems the presence of two dominating lcp sources when analysing the forward part of the c.m. as a function of impact parameter : intermediate velocity (IV) and projectile-like (PL) sources. The PL source velocity is evolving with centrality while the IV source is located at the c.m. velocity The gas of neutrons and protons in equilibrium with clusters has to be looked for in the IV source \cite{QinPRL108} since the other source is mostly producing lcp by secondary decays.
A simple and efficient way to minimize contribution from PL source is to apply a 60$^{0}$-90$^{0}$ c.m. polar angular selection \cite{BougaultPRC97}.\par
In the following the data concerning lcp are selected via: (i) central events, (ii) emission almost perpendicular to the beam direction in the c.m.(60$^{0}$-90$^{0}$ polar angular range).
\section{Expanding gas source and surface velocity}
\begin{figure}[ht]
\begin{center}
\resizebox{0.48\textwidth}{!}{%
   \includegraphics{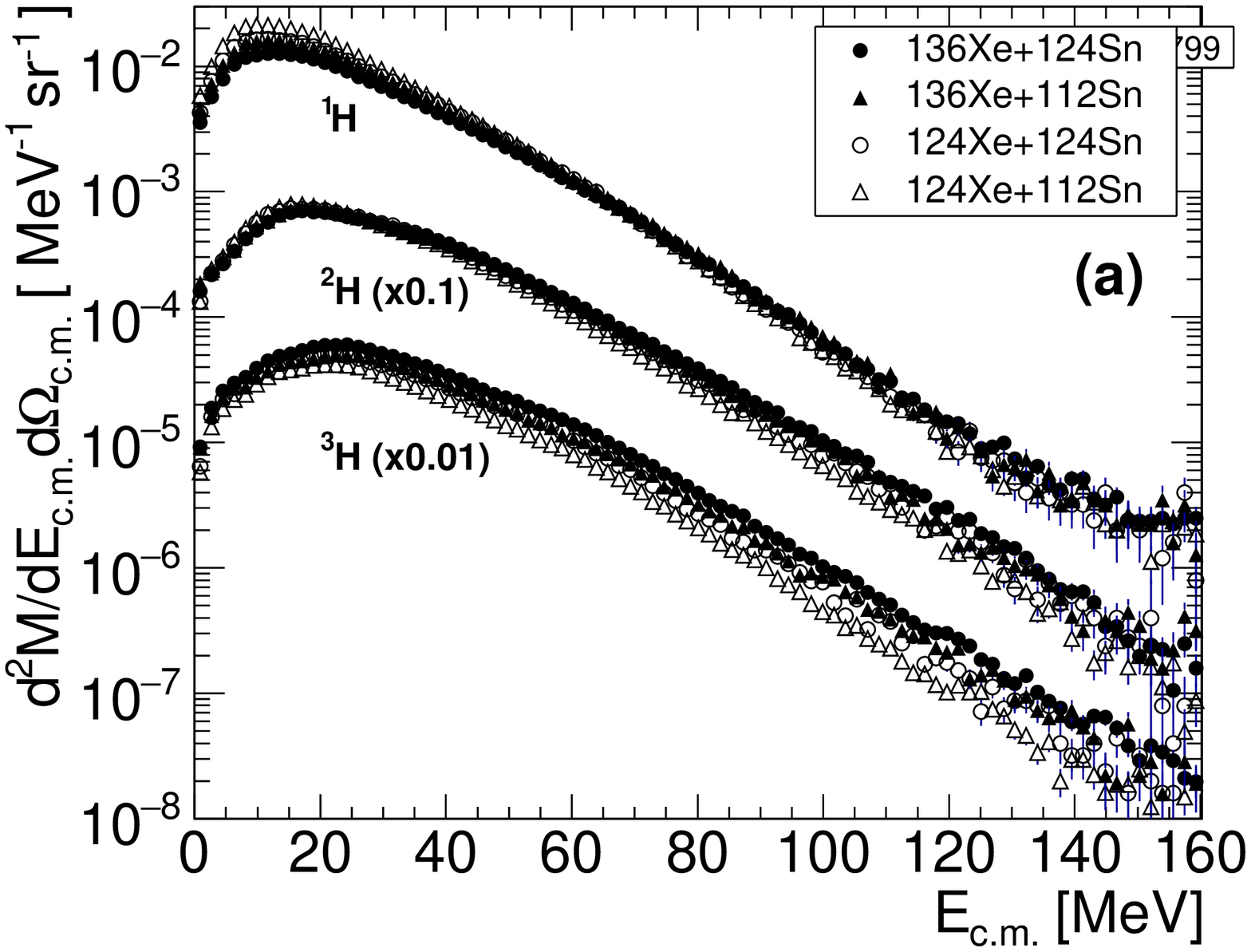}
   } 
\resizebox{0.48\textwidth}{!}{%
   \includegraphics{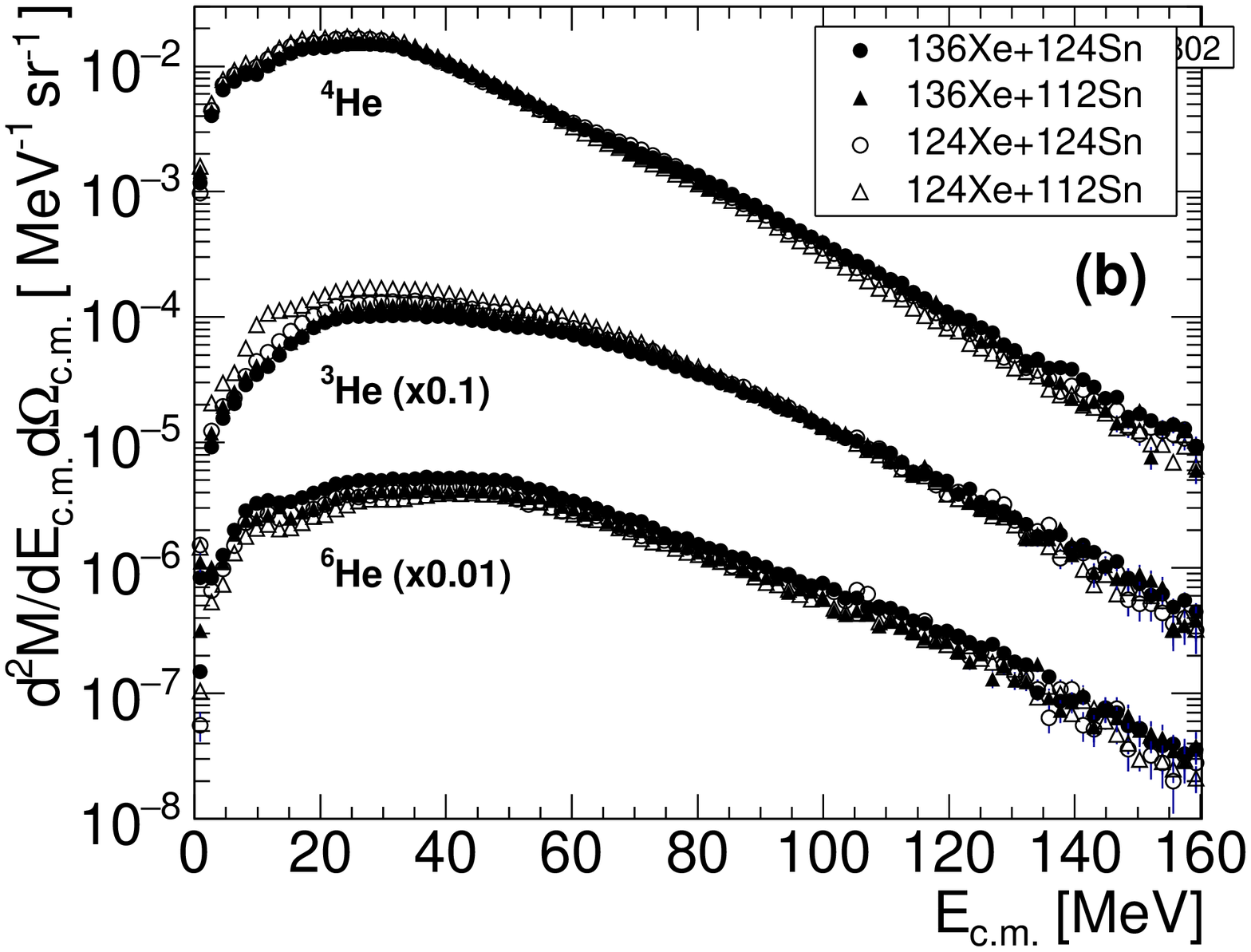}
   }         
\caption{Multiplicity of Z=1 (a) and Z=2 (b) isotopes as a function of their centre of mass energy for the four studied systems (60$^{0}$-90$^{0}$ polar angular range).}
\label{FigEcm}
\end{center}  
\end{figure}
The framework of the analysis assumes that the characteristics of the selected lcp correspond to the outcome of a whole chain of evolution of the IV gas source. Clusters are formed during the space-time evolution of the expanding gas of nucleons.
It is assumed that, for each time step, neutrons and protons are in equilibrium with clusters.
Equilibrium means that
although each collision is a continuously evolving dynamical process with no notion of temporal equilibrium, after a certain time in each collision the composition of the gas of nucleons and light clusters is frozen and corresponds to that which is detected, this is called (chemical) freeze-out. Because the measured properties 
\cite{BougaultPRC97} of the set of (sub-)events such as species multiplicity can be reproduced by assuming an unbiased population of the phase space available for the ensemble of freeze-out configurations produced by the collisions, then we may say that the observed (sub-)events are compatible with statistical equilibrium \cite{ColeISN1995} \cite{ChomazEPJA30} \cite{BorderiePPNP105}.\par 
The key observable is the velocity (v$_{surf}$) of the particles in the IV frame prior to acceleration
by the Coulomb field of the remaining charged material
\cite{AwesPRC24}. Calculations \cite{WangPRC72} indicate that the surface velocity decreases with increasing average emission time and therefore v$_{surf}$ values may be used to select different time steps of the gas part evolution, i.e. of the clusterisation process, fastest particles corresponding to earliest emission times.\par
A fitting procedure of the observed energy spectra is performed in \cite{QinPRL108} in order to disentangle the different lcp productions and to deduce the Coulomb repulsion on lcp caused by the remaining charged matter. We have adopted another method since the barrier parameters are generaly poorly defined in a multi-fit procedure \cite{HagelPRC62}.\par 
The characteristics of selected H and He elements are shown in figure \ref{FigEcm} through c.m. energy spectra for the four studied systems. 
The spectra are double differential multiplicities. They are c.m. energy spectra for particles detected in the 60$^{0}$-90$^{0}$ c.m. polar angular range normalized to the number of central events for each studied system, normalized to the c.m. energy bin size and normalized to the 60$^{0}$-90$^{0}$ c.m. solid angle. This is then the number of $^{1}$H, $^{2}$H, $^{3}$H, $^{3}$He, $^{4}$He, $^{6}$He per event, per MeV and per steradian.
The chemistry of lcp production is system dependent since neutron-rich/poor isotope productions are following the neutron-richness/poorness of the entrance channel systems.\par
In table \ref{TableEc} we present a calculation using the spectra of figure \ref{FigEcm}. 
Coulomb energy value per unit charge (E$_{C}$) is varied and 
we calculate the particle emission Coulomb boost caused by the remaining charge after complete emission of the particles of the expanding gas. The average charge of the expanding gas in the 60$^{0}$-90$^{0}$ c.m. angular range is the charge integrated over the multiplicity spectra for E$_{c.m.}$ greater than Z times E$_{C}$ for each particle and then summed together. This average charge approximated to the nearest integer is presented in the second column.
The numbers quoted in the third column are the remaining charge over 4$\pi$ after the complete emission of the particles of the expanding gas assuming an isotropic emission in the c.m.; this charge is responsible for the Coulomb boost. Since we are considering an expanding gas of nucleons and clusters at very low densities therefore the boost from the gas itself is negligible and we only consider the boost from the remaining part which is considered to be at much higher density.
The amount of the boost per atomic charge (fourth column) is then calculated by using standard Coulomb barrier formula for surface emission of H or He from the remaining charge. Comparing the values between first and last columns which should be identical, only the third hypothesis is consistent (E$_{C}$=10 MeV).
The deduced Coulomb boost may be considered as an average value since, event by event, the size of the remaining charge varies.
\par
\begin{table}[h]
\caption{\label{TableEc}$\Sigma$Z$_{lcp}$: sum of atomic number of particles having c.m. energy greater than Z times E$_{C}$ (hypothesis) in the c.m. considered angular range. 104-4$\Sigma$Z$_{lcp}$: 4$\pi$ remaining atomic number in relation to the total system charge minus total atomic number of the gas. Last column: calculated Coulomb barrier per Z between lcp and 104-4$\Sigma$Z$_{lcp}$.}
\begin{indented}
\item[]\begin{tabular}{@{}llll}
\br	
E$_{C}$&$\Sigma$Z$_{lcp}$&104-4$\Sigma$Z$_{lcp}$&E$_{C}$\\
hypothesis&[60$^{o}$,90$^{o}$]&4$\pi$&calculated\\
\mr	  
0 MeV&8&72&10 MeV\\	  
5 MeV&7&76&10 MeV\\
10 MeV&6&80&11 MeV\\
15 MeV&5&84&12 MeV\\		  	  	
20 MeV&3&92&13 MeV\\		  	  	  	    
\br
\end{tabular}
\end{indented}
\end{table}
The key observable (v$_{surf}$) of the gas part is then calculated taking into account a Coulomb Barrier of 10 MeV per atomic number with an error estimated to be $\pm$ 2 MeV.\par
\begin{figure}[ht]
\centering
\resizebox{0.48\textwidth}{!}{%
   \includegraphics{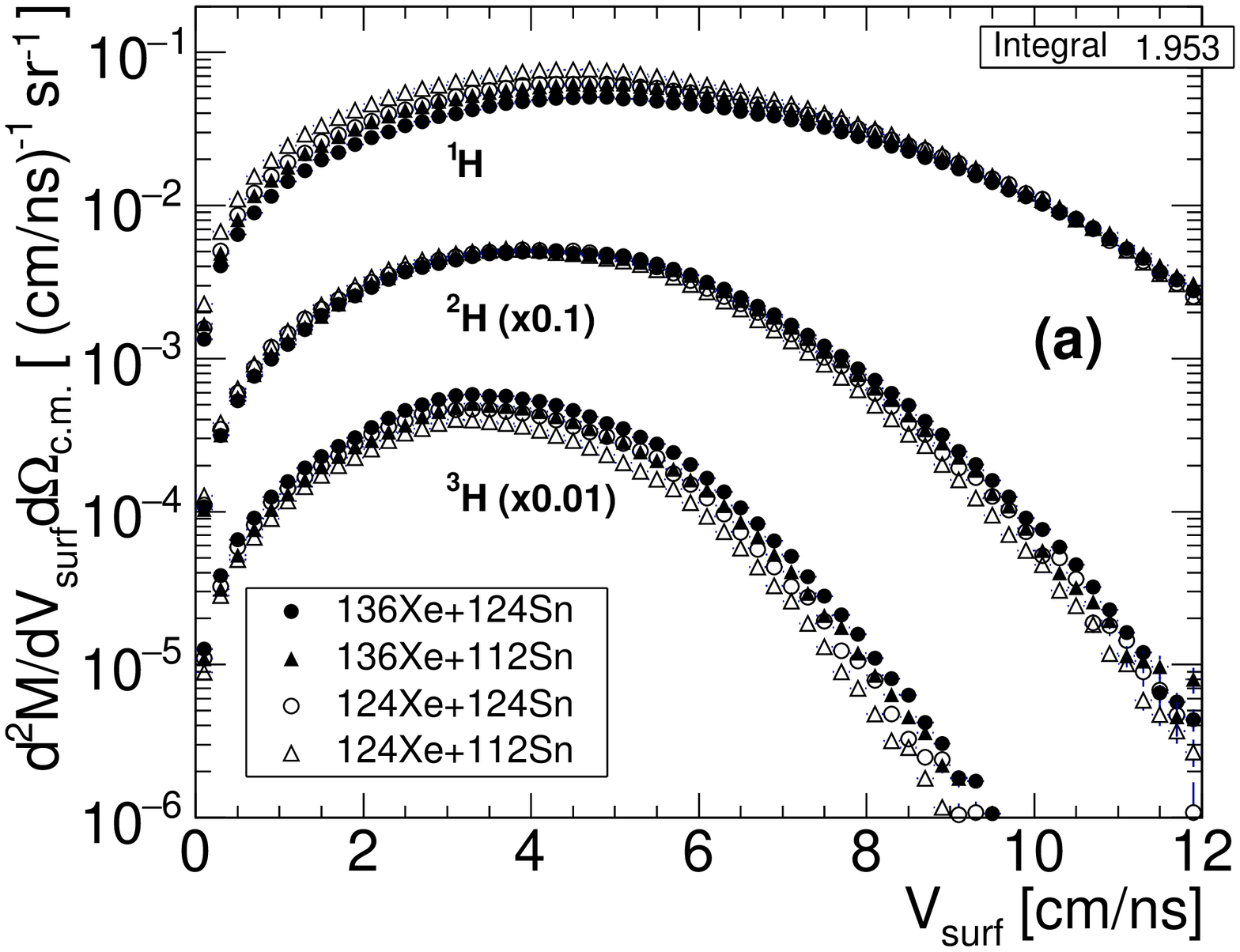}
   } 
\resizebox{0.48\textwidth}{!}{%
   \includegraphics{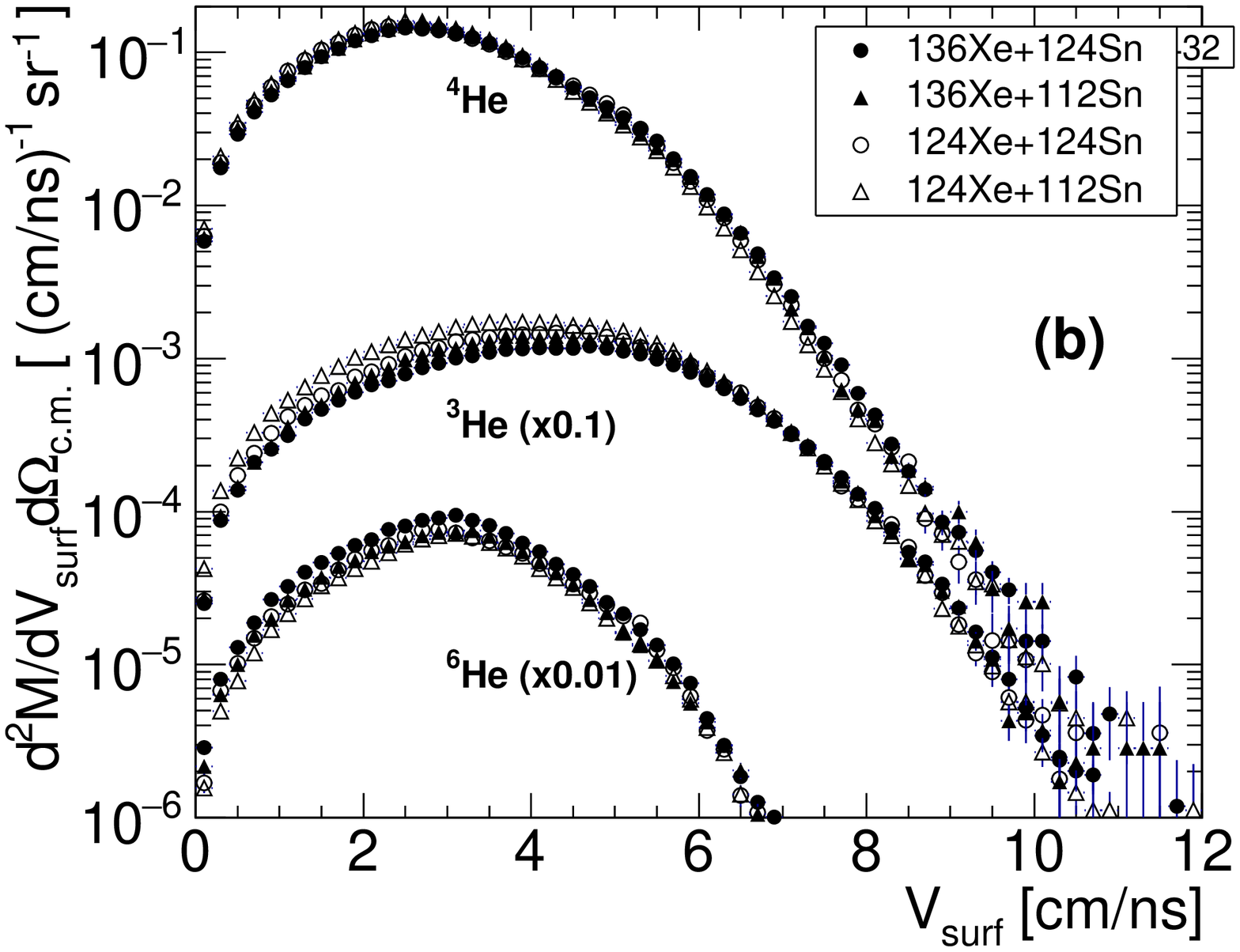}
   }         
   \caption{Multiplicity of Z=1 (a) and Z=2 (b) isotopes as a function of their reconstructed surface velocity for the four studied systems (60$^{0}$-90$^{0}$ polar angular range).}  
\label{FigVsurf}
\end{figure}

   
The sub-event selection is thus: (i) emission perpendicular to the beam direction in the c.m.(60$^{0}$-90$^{0}$ polar angular range), (ii) particles having c.m. energy greater than Z times 10 MeV.
The corresponding surface velocity spectra are presented in figure \ref{FigVsurf}, v$_{surf}$ being related to (E$_{c.m.}$-ZE$_{C}$)/A for a cluster (A, Z), with E$_{C}$=10 MeV. 
\section{Volume in momentum space}
The final-state coalescence model \cite{ButlerPRL7} proposed to explain deuteron formation was phenomenologically extended to heavier fragments \cite{ScPR129}. It is assumed that nucleons coalesce when located together inside a volume in momentum space of size $4\pi P_{0}^{3}/3$, with $P_{0}$ a parameter. To take into account for the Coulomb field generated by the remaining charge we use the Coulomb corrected coalescence model formalism of \cite{AwesPRC24} to determine coalescence parameter $P_{0}$ for each isotope as in \cite{QinPRL108}.\par
In the c.m. frame the relationship between the observed cluster and proton differential cross sections is
\vspace{12pt}
\begin{equation}
\label{EqCoalescence}
\fl
\frac{d^{2}M(A,Z)}{dE_{c.m.}d\Omega_{c.m.}} =R_{np}^{N}\frac{A^{-1}}{N!Z!}
\left(\frac{4\pi P_{0}^{3}}{3[2m^{3}(E^{proton}_{c.m.}-E_{C})]^{1/2}}\right)^{A-1}
\left(\frac{d^{2}M(1,1)}{dE^{proton}_{c.m.}d\Omega_{c.m.}}\right)^{A}
\end{equation}
\vspace{12pt}
where the cluster differential multiplicity, M(A,Z), having a Coulomb corrected energy $E_{c.m.}-ZE_{C}$ is related to the proton
differential multiplicity, M(1,1), at the same Coulomb corrected energy per nucleon, E$^{proton}_{c.m.}$-E$_{C}$.
E$_{C}$ is the Coulomb repulsion per unit charge, m is the nucleon mass, 
$(E_{c.m.}-ZE_{C})/A=E^{proton}_{c.m.}-E_{C}$ thus $E_{c.m.}=AE^{proton}_{c.m.}-NE_{C}$ for a nucleus composed of Z protons and N=A-Z neutrons and R$_{np}$ is the neutron to proton ratio. The neutron spectra is assumed to be identical to the proton spectra once the Coulomb correction has been applied.\par
Contrary to the original coalescence analysis \cite{NagamiyaPRC24} which determined an average value of P$_{0}$, here the goal is to characterize the evolution of the volume of the gas as a function of time, therefore of v$_{surf}$ or (E$_{c.m.}$-ZE$_{C}$)/A.\par
\begin{figure}[ht]
\centering
\resizebox{0.48\textwidth}{!}{%
   \includegraphics{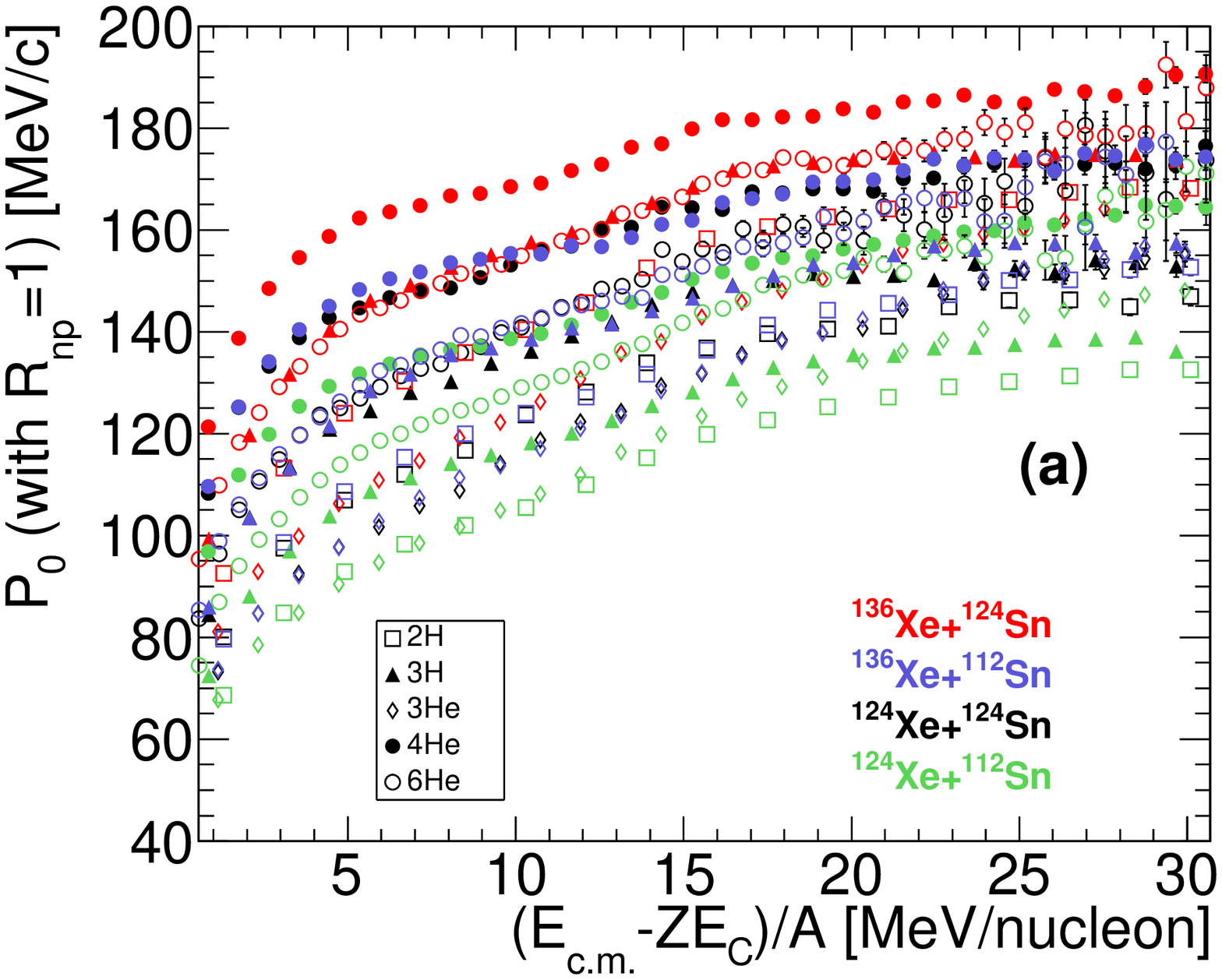}
   } 
\resizebox{0.48\textwidth}{!}{%
   \includegraphics{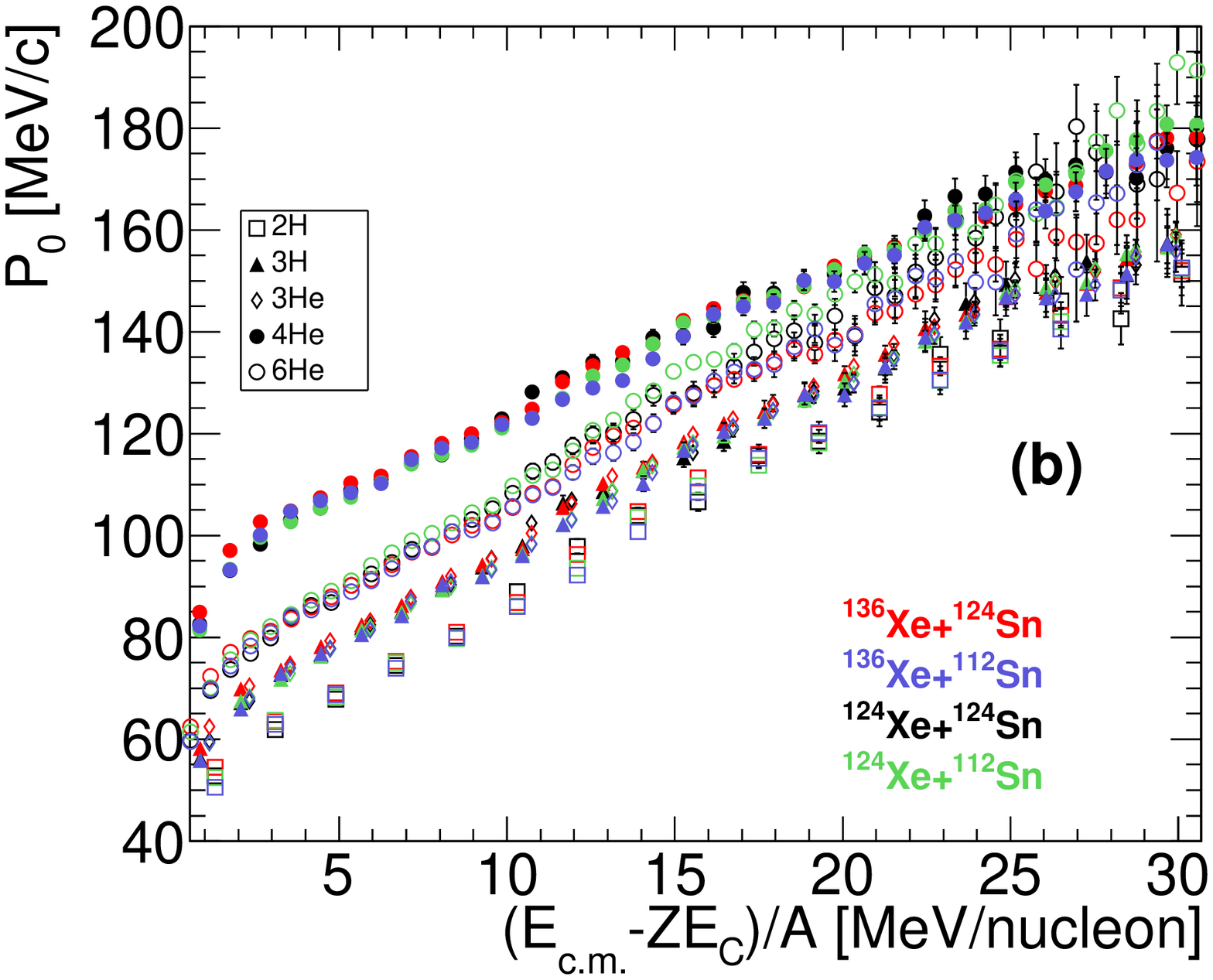}
   }           
   \caption{P$_{0}$ as a function of cluster energy per nucleon in the source frame before Coulomb acceleration. Results for the four systems and the five isotopes are shown. R$_{np}$=1 (a). R$_{np}$ from free neutrons and protons (b).}
\label{FigP0}  
\end{figure}

Neutrons are not measured in our experiment but hypotheses concerning the neutron to proton ratio may be tested by comparing results given by the four entrance channels ($^{136,124}$Xe+$^{124,112}$Sn) since P$_{0}$ should not depend on them. In a first step we used equation \ref{EqCoalescence} regardless of the neutron to proton ratio (R$_{np}$=1). The parameter P$_{0}$ as a function of (E$_{c.m.}$-ZE$_{C}$)/A is presented in figure \ref{FigP0}(a). We see that P$_{0}$ depends on cluster size and we observe large differences for the four different entrance channels.\par
Following the coalescence model \cite{CsernaiPR131} and as in \cite{QinPRL108}, the unmeasured neutron multiplicity at a given time, determined by v$_{surf}$ or (E$_{c.m.}$-ZE$_{C}$)/A, is given by the proton multiplicity multiplied by $^{3}$H/$^{3}$He multiplicity ratio \cite{FamianoPRL97} for that time. The question asked here is which proton to neutron ratio should be applied for each time step: the global ratio taking into account all species or the ratio calculated with only free nucleons. It turns out that the only ratio which preserves entrance channel independence is the ratio calculated with free neutrons and protons as shown in figure \ref{FigP0}(b). 
In the coalescence model the proton cross section used is the original production, before any composites are formed. 
The observed proton cross section, figure \ref{FigEcm}, is the original one minus the protons bound in the clusters.
This is also true for the neutron to proton ratio. The observed proton cross section is not the original one and figure \ref{FigP0} demonstrates the validity of using observed nucleon spectra. There is therefore a contradiction. It may be that by using the original nucleon spectra the appropriate neutron to proton ratio is also the original ratio, but this is impossible to prove experimentally since original proton energy spectra cannot be reconstructed.\par
The chemical equilibrium model \cite{MekjianPRC17} \cite{MekjianPLB89} also predicts a power law for the momentum space density of a composite nucleus relative to protons but is not affected by the previous arguments, since the law of mass action involves the free proton and neutron concentrations already reduced by composite formation.\par 
The results presented in this paragraph thus confirm the validity of the thermal model. A volume, V$_{0}$, which represents the spatial region (assumed spherical) over which chemical equilibrium is established will be calculated for each isotope using differential cross section power law\cite{MekjianPLB89} \cite{DasGuptaPR72}:
\vspace{12pt}
\begin{equation}\label{EqModeleEquilibre}
\fl
\frac{d^{3}M(A,Z)}{d^{3}p_{A}} =R_{np}^{N}\frac{(2s+1)~e^{B(A,Z)/T}}{2^{A}}\left(\frac{h^{3}}{V_{0}}\right)^{A-1}
\left(\frac{d^{3}M(1,1)}{d^{3}p}\right)^{A}
\end{equation}
\vspace{12pt}
where h is Planck's constant, s the spin of the cluster (Z, A, N=A-Z), B(A,Z) its ground state binding energy, T the temperature of the statistical ensemble describing the gas, $p$ the free proton momentum and $p_{A}=Ap$ the cluster momentum, both in the source frame before Coulomb acceleration ($p=mv_{surf}$).\par
The volume V$_{0}$ for each isotope (A,Z) is related to the volume in momentum space $4\pi P_{0}^{3}/3$ (P$_{0}$ of figure \ref{FigP0}(b)) by the relationship
\vspace{12pt}
\begin{equation}\label{EqVolume}
\fl
V_{0}=\frac{3h^{3}}{4\pi P_{o}^{3}}\left[(2s+1)\left(\frac{Z!N!A^{3}}{2^{A}}\right)e^{B(A,Z)/T}\right]
^{1/(A-1)}
\end{equation}
\vspace{12pt}
\section{Temperature and surface velocity confidence interval}
In this paragraph we will study the temperature evolution of the expanding gas source as a function of v$_{surf}$. 
\par  
The determination of the volume with equation \ref{EqModeleEquilibre} requires the measurement of the temperature. Following the statistical approach of \cite{AlbergoNCA89}, as in \cite{QinPRL108}, the temperature for a given v$_{surf}$ value is deduced from the multiplicities, M(A,Z), of $^{2}$H, $^{3}$H, $^{3}$He, $^{4}$He at the same v$_{surf}$ value with B(A,Z) the ground state binding energy of isotope (A,Z)
\vspace{12pt}
\begin{equation}\label{EqT}
\fl
T =\frac{B(4,2)+B(2,1)-B(3,2)-B(3,1)}{\ln(\sqrt{9/8}(1.59~R_{v_{surf}}))} MeV
~{\rm with}~R_{v_{surf}}=\frac{M(2,1)M(4,2)}{M(3,1)M(3,2)}
\end{equation}
\vspace{12pt}
This relationship, valid for particles emitted from a single source at temperature T and having a Maxwellian spectrum, is applied for each bin of v$_{surf}$.\par
\begin{figure}[ht]
\centering
\resizebox{0.48\textwidth}{!}{%
   \includegraphics{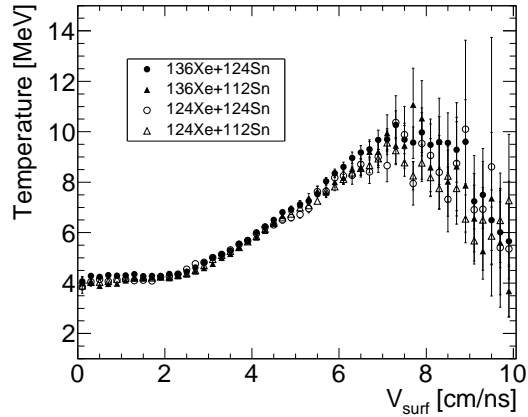}
   }         
   \caption{Temperature versus surface velocity for the four systems.}
   \label{FigTemperature}
\end{figure}

The deduced temperature values as a function of v$_{surf}$ are shown in figure \ref{FigTemperature}. Temperatures are independent of the entrance channel, $^{136,124}$Xe+$^{124,112}$Sn, as expected since the bombarding energy is identical. This result was not self-evident because the multiplicities, especially $^{3}$H and $^{3}$He, are very system dependent (see figure \ref{FigVsurf}). 
For V$_{surf}$ below 6.5 cm/ns, $^{2}$H and $^{4}$He productions are entrance channel system independent, the spectra are the same for the four studied systems. It is not the case for $^{3}$H and $^{3}$He productions; as expected $^{3}$H production is growing with neutron richness of the system while $^{3}$He is decreasing.
The temperature depends on $^{4}$He, $^{2}$H, $^{3}$He and $^{3}$H productions. Since only $^{3}$H and $^{3}$He productions depend on the entrance channel N/Z, therefore in order for the temperature to be identical for the four studied systems $^{3}$H and $^{3}$He multiplicities must counterbalance each other almost exactly.
This result further strengthens the validity of the equilibrium model for describing the data. For higher v$_{surf}$ values the error bars prevent us from drawing definite conclusions.\par   
The evolution of T as a function of v$_{surf}$ is similar to that presented in Figure 3 of \cite{WangPRC72} and figure 1 in \cite{KowalskiPRC75}. In our case the temperature is constant for v$_{surf}$ below 3 cm/ns, then it increases to reach a maximum around 9 MeV for v$_{surf}$ about 6.5 cm/ns. Then it seems to decrease with increasing v$_{surf}$. Figure \ref{FigTemperature} indicates that the scenario of the IV source which cools with time is valid for v$_{surf}$ greater than 3 cm/ns and lower than 6.5 cm/ns. The high limit is adopted for lack of cluster population while below 3 cm/ns several lcp production mechanisms are probably present.\par
In view of the results mentioned in this paragraph, the surface velocity interval between 3 and 6.5 cm/ns represents the confidence interval for the subsequent analysis.\par 
\section{Volume from thermal model}
In this paragraph we will study the volume evolution of the expanding gas source as a function of v$_{surf}$ using equation \ref{EqModeleEquilibre} and \ref{EqT}.\par
\begin{figure}[ht]
\centering
\resizebox{0.48\textwidth}{!}{%
   \includegraphics{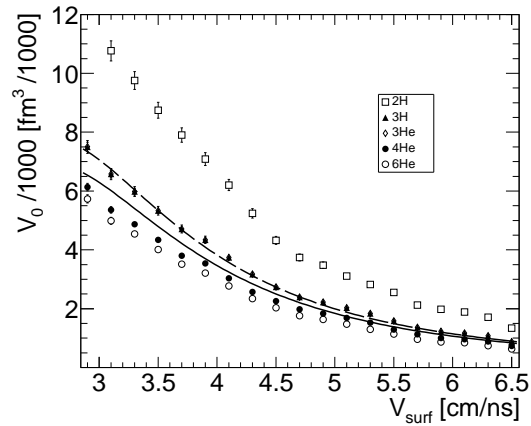}
   }         
   \caption{Volume from thermal model as a function of surface velocity for the five isotopes ($^{124}$Xe+$^{124}$Sn). The lines represent two fits of the average volumes of A$\ge$3. The full line is a fit of data using the ratio of free neutrons to free protons with equation \ref{EqNsurZfree} whereas the dashed line is a fit of data without the temperature dependence of free neutrons to free protons ratio.}
   \label{FigVolume}
\end{figure}
The results are presented in figure \ref{FigVolume}. The gas source volumes are increasing with decreasing v$_{surf}$. This observation is compatible with the picture of an evolving expanding source with time.
There exists a dependence on the composite particle which has been interpreted in terms of the finite size of the composite \cite{DasGuptaPR72}.\par
In the coalescence model the free neutron to free proton ratio is given by the $^{3}$H/$^{3}$He multiplicity ratio. In the equilibrium model \cite{AlbergoNCA89} this ratio (R$_{np}$ of equation \ref{EqModeleEquilibre}) is given by
\vspace{12pt}
\begin{equation}\label{EqNsurZfree}
\fl
R_{np}=\frac{M(3,1)}{M(3,2)}~e^{((B(3,2)-B(3,1))/T)}
\end{equation}
\vspace{12pt}
where T is the temperature and B(A,Z) the ground state binding energy of isotope (A,Z).\par
The volume evolution of the expanding gas source as a function of v$_{surf}$ is calculated using equation \ref{EqNsurZfree} for R$_{np}$. Without this temperature dependence contained in the exponential term of equation \ref{EqNsurZfree} the deduced volumes of $^{3}$H and $^{3}$He are different. The validity of introducing this temperature dependence is confirmed by Relativistic Mean-Field approach calculations \cite{PaisPRC97} where deviations of about 20\% concerning the free neutron proton ratio can occur for temperatures around 4 MeV if the exponential term of equation \ref{EqNsurZfree} is not introduced \cite{PaisPrivate}. This dependence on temperature is a difference as compared to \cite{QinPRL108}.\par
The volumes extracted for deuterons are larger as in \cite{QinPRL108}. This could be explained by the fragility of the deuteron and its survival probability once formed \cite{QinPRL108}. Therefore, as in \cite{QinPRL108}, the average volumes over which chemical equilibrium is established are derived from A$\ge$3 clusters. The average volumes are represented by the full line in figure \ref{FigVolume}. The full line is the result of a fitting procedure of the average volumes using a Landau function for each studied system. The dashed line is the Landau fit of average volumes without the temperature dependence of R$_{np}$. We see that the temperature dependence of equation \ref{EqNsurZfree} is small for high v$_{surf}$ and becomes important for low v$_{surf}$ where the temperature is the lowest.\par 
\section{Density}
In this paragraph we will study the density evolution of the expanding gas source as a function of v$_{surf}$.\par
The average volumes of figure \ref{FigVolume} are free volumes since they were calculated in a framework of point-like clusters. In order to deduce the total average volumes it is necessary to add the excluded volumes represented by the volumes occupied by the clusters themselves. The density is related to total average volume.\par
The evolving source volume for each bin of v$_{surf}$, for a given time t, is
\vspace{12pt}
\begin{equation}\label{EqVolumeTotal}
\fl
V_{t} = V_{0}+\frac{4}{3}\pi r_{0}^{3}A_{t}
\end{equation}
\vspace{12pt}
where V$_{t}$ is the total volume of the gas at time t, V$_{0}$ is the volume deduced from the thermal model at time t, A$_{t}$ is the total mass of the gas at time t and r$_{0}$ is the average particle radius (1.3 fm). 
The density at time t is $\rho_{t}=A_{t}/V_{t}$.\par
\begin{figure}[ht]
\centering
\resizebox{0.6\textwidth}{!}{%
   \includegraphics{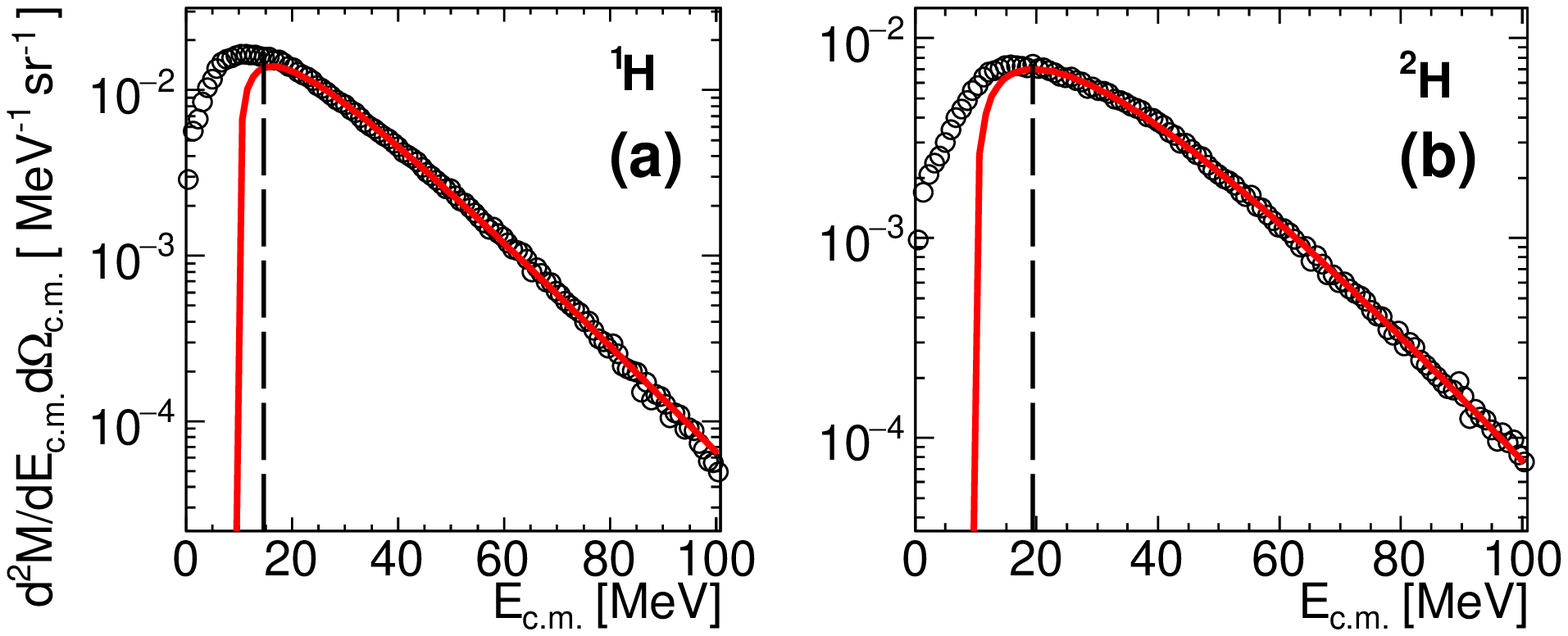}
   } 
\resizebox{0.6\textwidth}{!}{%
   \includegraphics{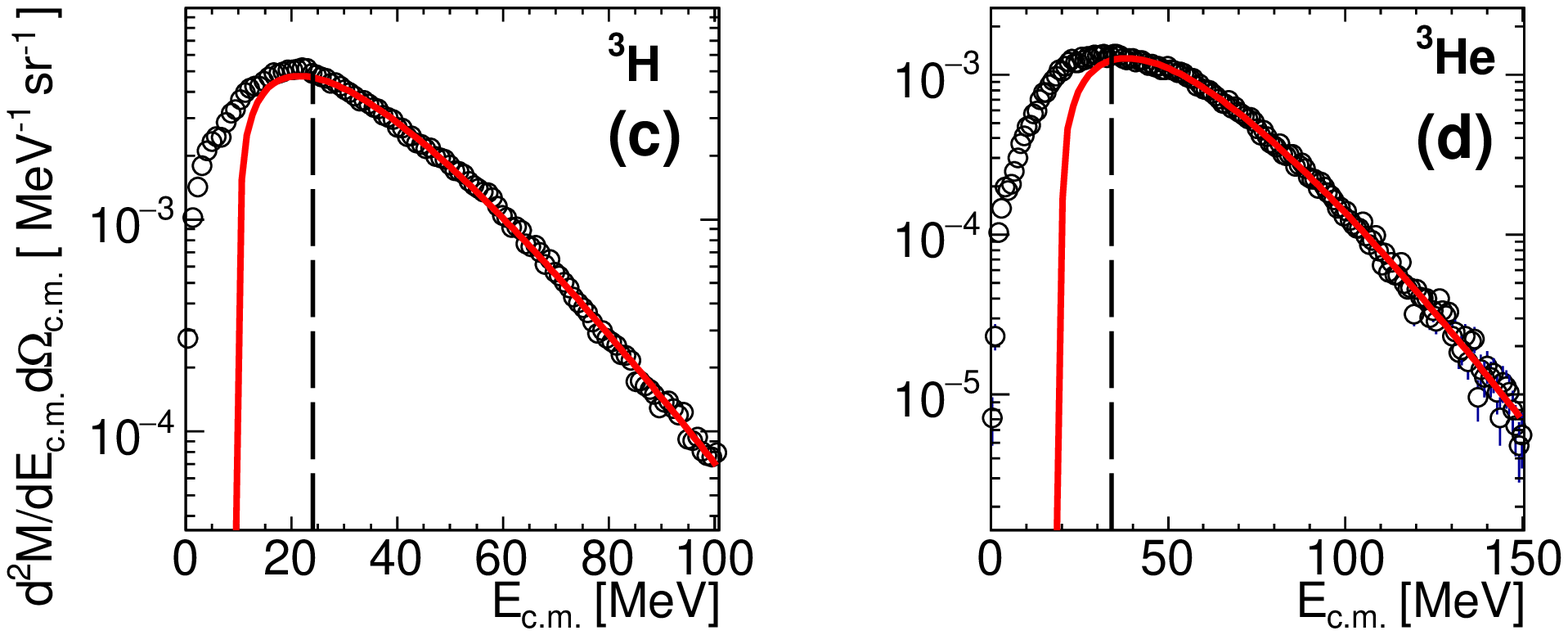}
   } 
\resizebox{0.6\textwidth}{!}{%
   \includegraphics{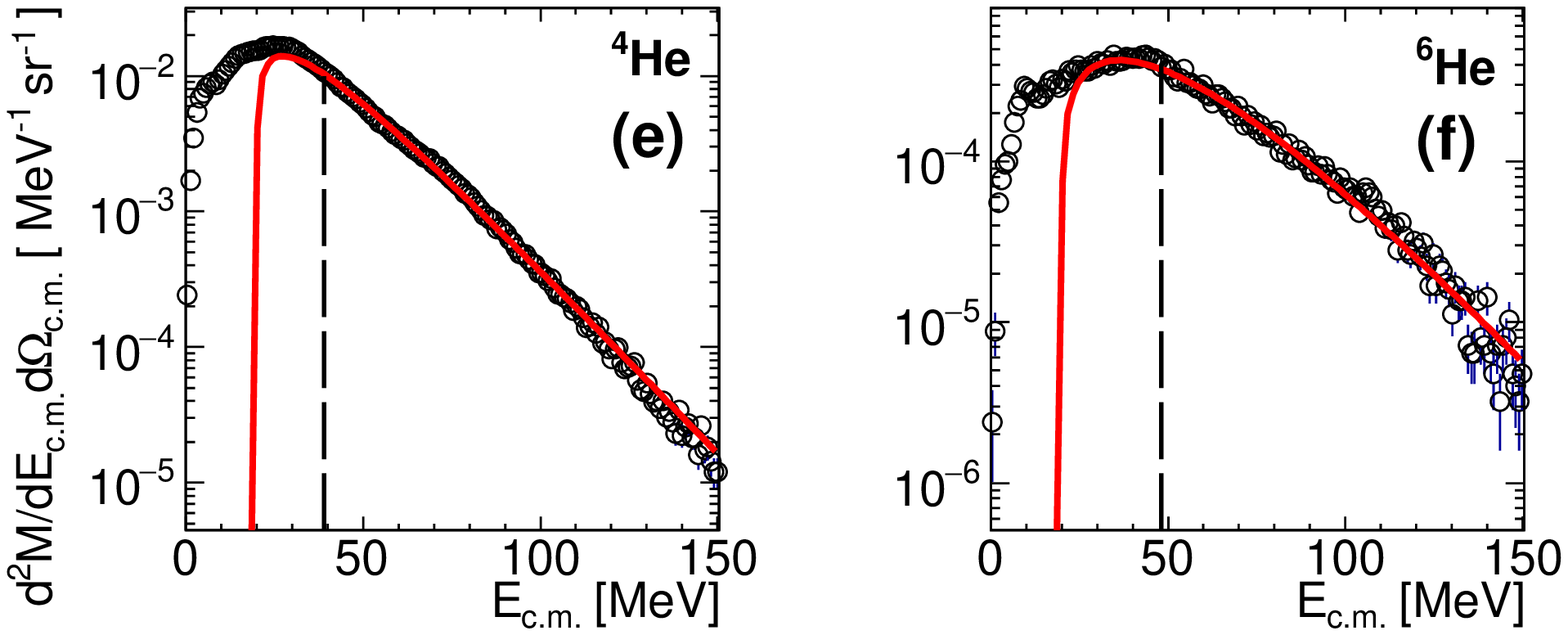}
   }           
   \caption{Fits of E$_{cm}$ isotope spectra: $^{1}$H (a), $^{2}$H (b), $^{3}$H (c), $^{3}$He (d), $^{4}$He (e), $^{6}$He (f). Full lines are the fits performed independently for each isotope, the dashed lines correspond to v$_{surf}$=3 cm/ns. $^{124}$Xe+$^{124}$Sn system.}
   \label{FigFitsEcm}
\end{figure}
The knowledge of A$_{t}$ at time t is related to the knowledge of A$_{source}$, the mass of the gas source at the beginning of the expansion process, through $A_{t}=A_{source}-A_{exp}$, where A$_{exp}$ is the mass expelled before time t. In other words, for a given v$_{surf}$, A$_{t}$ is the total mass integrated from 0 to v$_{surf}$.\par
It has been shown that the v$_{surf}$ interval between 0 and 3 cm/ns does not match with the picture of a cooling while expanding gas (confidence interval). Therefore we cannot rely directly on the data in this velocity range to calculate A$_{t}$. A$_{source}$ has thus been evaluated using a fitting procedure of E$_{cm}$ spectra of figure \ref{FigEcm}.\par     
The spectra were fitted independently by an expanding Boltzmann distribution \cite{BondorfNPA296}
\vspace{12pt}
\begin{equation}\label{EqFit}
\fl
\frac{d^{2}M(A,Z)}{dE_{c.m.}d\Omega_{c.m.}} =C~e^{-(E_{c.m.}-ZE_{C}+E_{Reff})/T_{eff}}
~sinh(2\sqrt{(E_{c.m.}-ZE_{C})E_{Reff}}/T_{eff})
\end{equation}
\vspace{12pt}
where E$_{C}$ is the Coulomb repulsion per unit charge, E$_{Reff}$ is related to the radial expansion energy of the source, T$_{eff}$ is the temperature and C a normalization factor.\par
The fits are performed for E$_{cm}$ values corresponding to v$_{surf}$ greater than 3 cm/ns and E$_{C}$ is set at 10 MeV, the other three parameters being left free. E$_{Reff}$ and T$_{eff}$ are effective parameters since they represent an integration over the whole expansion process.\par
\begin{figure}[ht]
\centering
\resizebox{0.48\textwidth}{!}{%
   \includegraphics{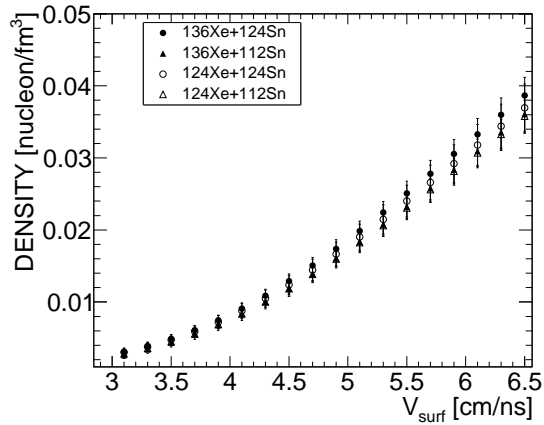}
   }         
   \caption{Density versus surface velocity for the four studied systems.}
   \label{FigDensity}
\end{figure}
The results of the fitting procedure are given in figure \ref{FigFitsEcm} for the $^{124}$Xe+$^{124}$Sn system.
We see that the spectra are quite well reproduced for the v$_{surf}$ confidence interval whose lower limit is represented by the dashed line. 
It should be mentioned that the proton spectrum is not so well reproduced between v$_{surf}$=3 and 4 cm/ns, which would tend to decrease the confidence interval. However, the latter has not been modified. 
Results for the three other studied systems are similar.\par
The gas source size, integrated over 4$\pi$, is calculated using fit integration. v$_{surf}>0$ neutron multiplicity is deduced from equation \ref{EqNsurZfree} and as far as their contribution to the source is concerned, they are assigned the percentage of protons. The gas source size (A$_{source}$) is $51\pm2.2$, $53\pm1.6$, $49\pm1.6$ and $51\pm1.5$ for $^{124}$Xe+$^{124}$Sn, $^{136}$Xe+$^{124}$Sn, $^{124}$Xe+$^{112}$Sn, $^{136}$Xe+$^{112}$Sn respectively. 
With regard to the confidence interval defined above (v$_{surf}$ values between 3 cm/ns and 6.5 cm/ns), as the mass of the evolving source (A$_{t}$) for a given v$_{surf}$ is taken as the total integrated mass between 0 and v$_{surf}$, it appears that the integrated mass between 6.5 cm/ns and infinity will have no influence on the following results.
\par
The density values ($\rho_{t}=A_{t}/V_{t}$) are presented in figure \ref{FigDensity} as a function of surface velocity in the confidence interval. 
Density is increasing with surface velocity, i.e. is decreasing with time. As in \cite{QinPRL108} very low values are achieved.\par
Concerning the 4$\pi$ isotropic distribution of the gas source, we should note that this requirement could be restricted to the 60$^{o}$-90$^{o}$ angular range ($\pi$) without changing density values, the source size being divided by 4 as well as the volume.
\section{Gas source characteristics}
In this paragraph we present the characteristics of the gas source during its evolution.\par
The gas source characteristics for each time step are given by its temperature, its composition and its density.
\begin{figure}[ht]
\centering
\resizebox{0.48\textwidth}{!}{%
   \includegraphics{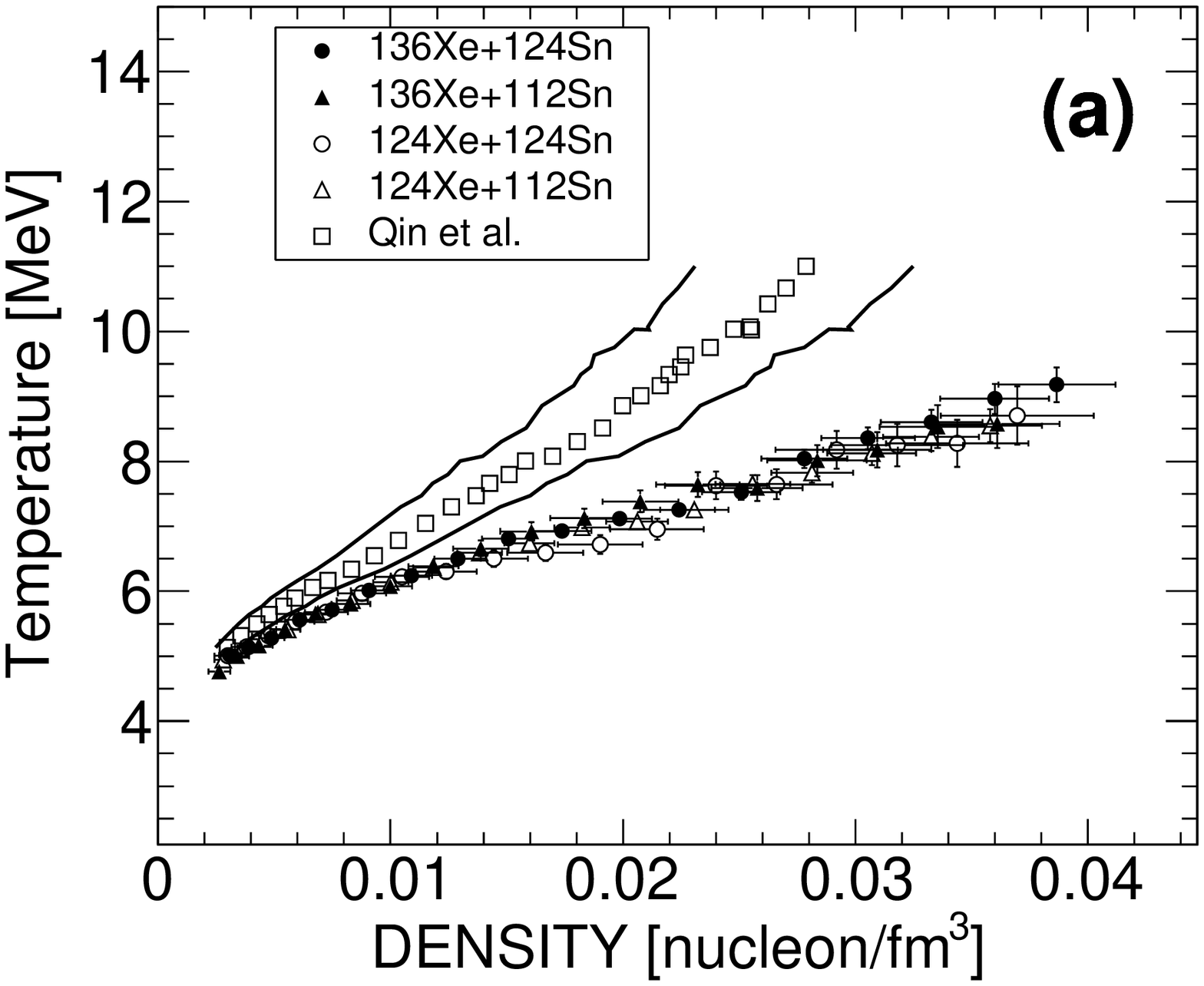}
   }
\resizebox{0.48\textwidth}{!}{%
   \includegraphics{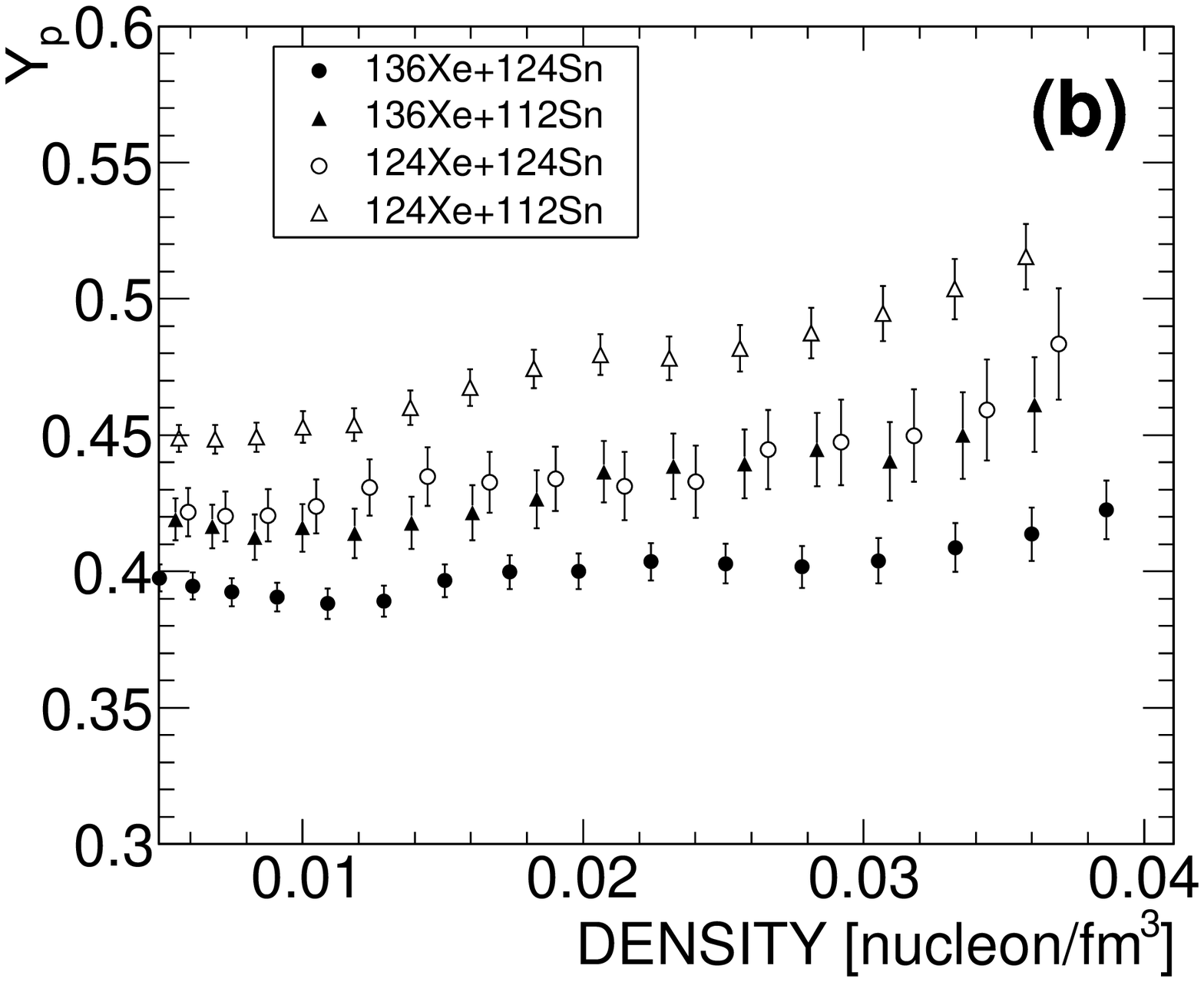}
   }             
\caption{Gas source characteristics. (a): temperature versus density for the four studied systems, squares are results from \cite{QinPRL108} with uncertainties represented by the lines.
(b): proton fraction as a function of density for the four studied systems; for clarity errors on density are not drawn. Density errors are indicated in figures \ref{FigDensity} and \ref{FigTvsDensity}(a).}  
   \label{FigTvsDensity}
\end{figure}
In fact, the value of the temperature and the composition of the evolving source for each time step is given by the temperature and the composition calculated with lcp multiplicities contained in each bin of v$_{surf}$. Only the calculation of density requires the knowledge of the total mass of the source for each time step (A$_{t}$). The temperature is calculated with equation \ref{EqT} and the neutron multiplicity from the proton multiplicity through equation \ref{EqNsurZfree}.\par
The density dependence of the temperature is given in figure \ref{FigTvsDensity}(a) for the four studied systems. The temperature, ranging from 5 to 9 MeV, does not depend on the entrance channel. The figure presents the thermodynamical path of the cooling of the source. The INDRA data are confronted to results from \cite{QinPRL108} obtained at higher bombarding energy (47A MeV). This difference in beam energy may explain the gap between the two thermodynamical paths. For very low densities, the temperatures are closer. We may also attribute these differences to different collision dynamical paths and therefore different characteristics for the expanding gas present at mid-rapidity since in our case the entrance channel system is quasi-symmetric (Xe+Sn) while in the case of Qin et al. it is asymmetric (Ar+Sn and Zn+Sn).\par  
The composition of the evolving source is given by the proton fraction, Z/A, whose density dependence is shown in figure \ref{FigTvsDensity}(b). The proton fraction values depend on the projectile plus target proton fractions, we note that they are identical for $^{136}$Xe+$^{112}$Sn and $^{124}$Xe+$^{124}$Sn systems in agreement with chemical equilibrium hypothesis.
The entrance channel proton fraction values are reached for low densities while there appears to be a trend towards larger proton fraction values as the density increases. This trend is larger for proton rich entrance channel systems.
This density dependence is essentially due to the $^{4}$He contribution to the global proton fraction. The $^{4}$He contribution evolves from 0.1 to 0.5 as the density decreases for the four studied systems, $^{4}$He clusters being more abundant for low densities. Therefore the global proton fraction is decreasing with decreasing density.\par  
\section{Equilibrium constants}
In this paragraph we will compare the INDRA equilibrium constants to the data of Qin et al. \cite{QinPRL108}.\par
\begin{figure}[ht]
\centering
\resizebox{0.48\textwidth}{!}{%
   \includegraphics{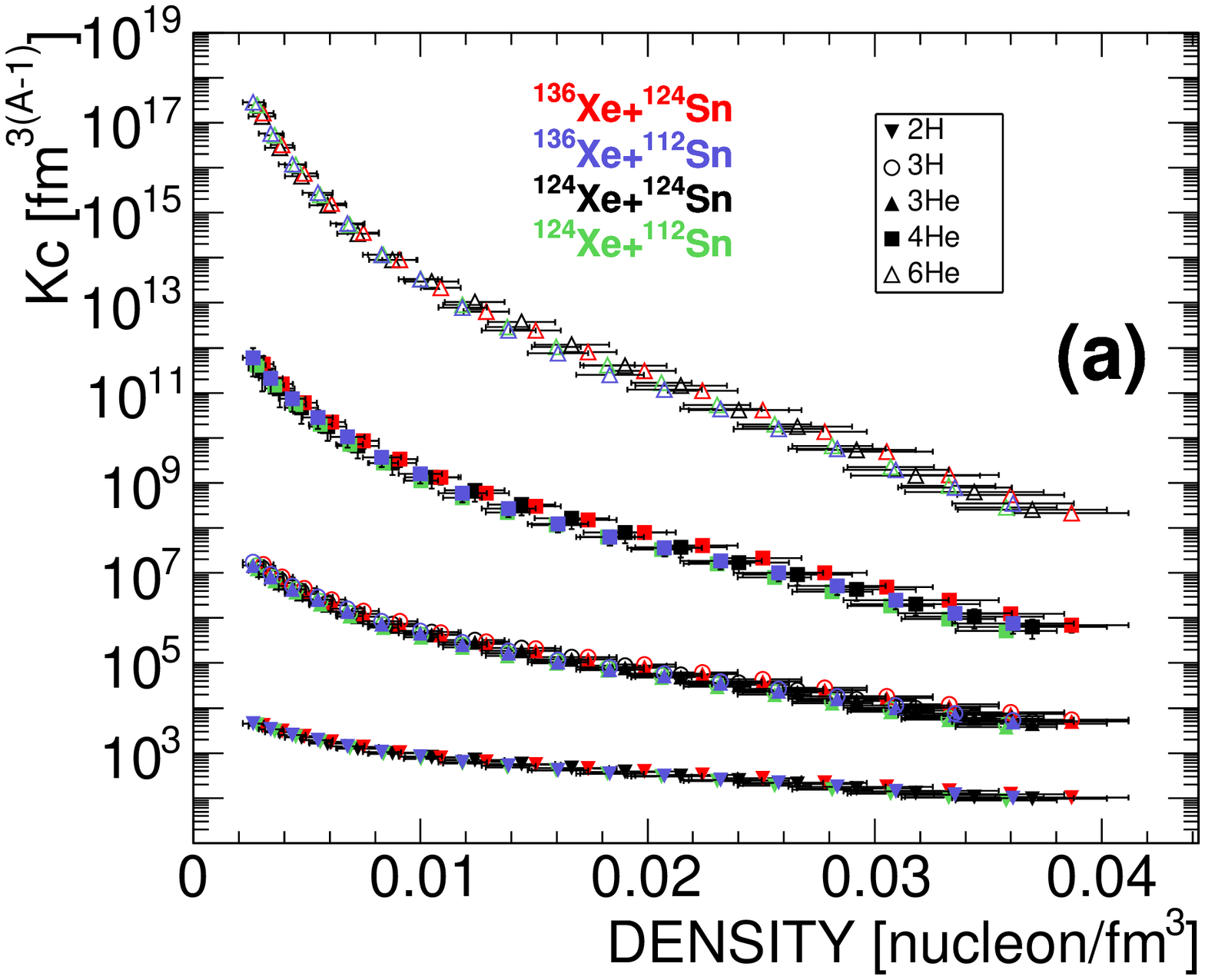}
   }
\resizebox{0.48\textwidth}{!}{%
   \includegraphics{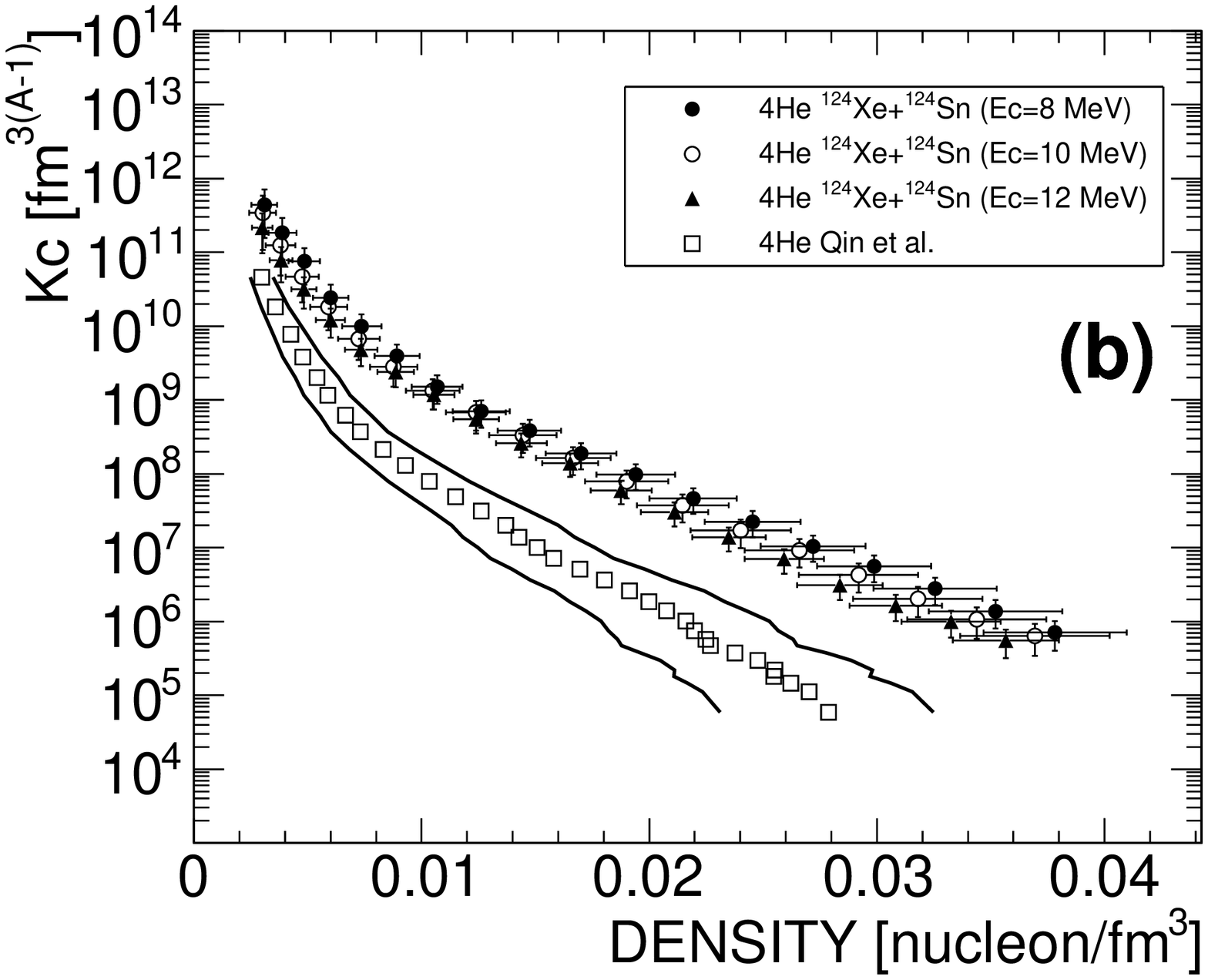}
   }            
   \caption{Equilibrium constants as a function of density. (a): for the studied isotopes and for the four studied systems. (b): $^{4}$He Equilibrium constant as a function of density, three results are shown for the $^{124}$Xe+$^{124}$Sn data to take into account for the error on the Coulomb barrier used to calculate v$_{surf}$, the squares are the Qin et al. \cite{QinPRL108} data with the uncertainties represented by the lines.}
   \label{FigKcvsDensity}
\end{figure}

Equilibrium constant for a cluster of mass A and atomic number Z is defined as
\vspace{12pt}
\begin{equation}\label{EqKc1}
\fl
K_{c}(A,Z)=\frac{\rho_{pa}(A,Z)}{\rho_{pa}(1,1)^{Z}~\rho_{pa}(1,0)^{N}}
\end{equation}
\vspace{12pt}
where $\rho_{pa}(A,Z)$ is the (A,Z) particle partial density ($\rho_{t}=\sum_{i} \rho_{pa}(A_{i},Z_{i})$). K$_{c}$ should depend only on density and temperature of the equilibrated gas source which makes it a universal parameter. The isotope equilibrium constants are shown in figure \ref{FigKcvsDensity}(a) for the four studied systems as a function of the gas source density ($\rho_{t}$).\par 
K$_{c}$ is easily reformulated as a function of mass fractions $w(A,Z)$
\vspace{12pt}
\begin{equation}\label{EqKc2}
\fl
K_{c}(A,Z) = \frac{w(A,Z)}{w(1,1)^{Z}~w(1,0)^{(A-Z)}}~\rho_{t}^{-(A-1)}
~{\rm with}~w(A,Z)=\frac{A~M(A,Z)}{\sum_{i}{A_{i}~M(A_{i},Z_{i})}}
\end{equation}
\vspace{12pt}
M(A,Z) being the multiplicity of particle (A,Z) calculated in a given v$_{surf}$ bin.\par
This to say that K$_{c}$ has an explicit dependence on $\rho_{t}^{-(A-1)}$ which explains its global density behaviour.
From figure \ref{FigKcvsDensity} we note that K$_{c}$ is independent of A/Z entrance channel within the error bars.\par
In figure \ref{FigKcvsDensity}(b), K$_{c}$ density dependence is presented for $^{4}$He. In order to take into account for the error on the Coulomb barrier per atomic number used to calculate v$_{surf}$ (E$_{C}=10\pm 2$ MeV) we present three results (E$_{C}=8, 10~$and$~12$ MeV) whose superposition indicates the final error on our data.\par
In figure \ref{FigKcvsDensity}(b) our data is also compared to results of Qin et al. \cite{QinPRL108}. Since the thermodynamical paths (figure \ref{FigTvsDensity}(a)) are not the same, we do not expect the two data sets to overlap. This is indeed the case. At very low densities, where the temperatures of the two experiments are close, we could have expected a better agreement, however, as we indicated earlier, the confidence interval below v$_{surf}=4~cm/ns$ (density lower than 0.008 nucleon/fm$^{3}$) is questionable for our data. Also we remind the reader that Qin et al. did not use equation \ref{EqNsurZfree} whose impact is important at low temperatures. Therefore there is no contradiction between the two results. However the full compatibility test has to be done with a model comparison.\par 
\section{The problem of binding energies}
In this paragraph we will comment the method used to extract the equilibrium constants and the density values.\par
Model comparisons with \cite{QinPRL108} data have been made \cite{HagelEurPhys50} \cite{HempelPRC91} \cite{PaisPRC97} and all the studies conclude that the measured equilibrium constants are not reproduced by an ideal gas modelisation. 
Medium effects are present and the consequence is the existence of a shift of the cluster binding energies \cite{RopkeNPA867} \cite{RopkePRC92} \cite{TypelPRC81} for $\rho > 0$.
The decrease in the binding energy arises from the indistinguishability between the nucleons inside the
clusters and the free nucleons (Pauli blocking). 
The procedure ignores the reduction in the binding energy of the clusters as
the density of the surrounding medium increases because it assumes values of the cluster binding energy in vacuum
in equations \ref{EqModeleEquilibre}, \ref{EqVolume}, \ref{EqT} and \ref{EqNsurZfree}.
\par
In summary, chemical constants measured by Qin et al. were compared to statistical models employing different treatments of the nucleon-nucleon and nucleon-cluster interactions with the purpose of constraining the expected cluster binding energy shifts in dense matter. Formulae used to extract the measured quantities, temperature and volume and free neutron number, explicitly assume that the cluster abundancies are uniquely governed by their vacuum binding energies, which is in contradiction with the purpose of the analysis. Furthermore in reference \cite{HempelPRC91} it is shown that the Qin et al. experimental points are not reproduced by an ideal gas hypothesis although all the experimental analysis is based supposing an ideal gas of classical clusters in thermodynamic equilibrium at temperature T in the grand-canonical ensemble. Indeed if in medium corrections were negligible, the measured chemical constants would agree with the ideal gas prediction.
There is therefore a contradiction in the method. This point has to be clarified.\par
\section{Conclusions}
Equilibrium constants as a function of density has been presented in the framework of the study made by Qin et al. \cite{QinPRL108}. 
A subset of events has been selected whose characteristics are consistent with a chemically- and thermally-equilibrated expanding gas of nucleons and clusters. The power law for the momentum space density of composite nuclei relative to protons has been used to extract the source volumes. It turned out that the chemical equilibrium model is more adapted to our data as compared to the coalescence model. The temperature values of the expanding source have been deduced using nuclear statistical equilibrium framework and entrance channel independence ($^{136,124}$Xe+$^{124,112}$Sn) of the result confirmed the validity of its use and allowed us to define a confidence interval for the subsequent analysis. The volume values of the expanding source have been extracted and 
an improvement, confirmed by theorical calculations, has been applied concerning the relationship between neutron and proton multiplicities as compared to the original work of Qin et al. Those volume values have been transformed in density values by determining the total source size. The thermodynamical path covered by the expanding source in the temperature-density plane is different to that of Qin et al. This is not abnormal because the bombarding energies of the two experiments are not the same. Equilibrium constant values as a function of density were presented for $^{2}$H, $^{3}$H, $^{3}$He, $^{4}$He, $^{6}$He and the result is not in contradiction with the results of Qin et al.\par
Finally, we raised the problem of the values of cluster binding energies. Indeed, the values used in the analysis are those in vacuum (zero density) while all comparisons with the models indicate the existence of a shift towards lower values and existence of in-medium effects. This point has to be clarified. 

\end{document}